\documentclass[sigconf,nonacm]{acmart}
\usepackage{subcaption}
\usepackage{fontawesome}
\usepackage{tabularx}
\usepackage{tikz}
\usetikzlibrary{arrows.meta,positioning,calc,fit,shapes.geometric,shapes.misc,backgrounds,shapes.symbols}
\tikzset{shield/.style={rounded corners, very thick, draw=red, fill=yellow!20}}

\definecolor{navy}{HTML}{274060}       
\definecolor{rose}{HTML}{F2D7D5}       
\definecolor{redc}{HTML}{C0392B}       
\definecolor{bluegray}{HTML}{E8F1F5}   
\definecolor{sky}{HTML}{5DADE2}        
\definecolor{forest}{HTML}{27AE60}     
\definecolor{charcoal}{HTML}{2C3E50}   

\tikzset{
  zoneTrusted/.style={fill=bluegray, draw=sky, rounded corners=6pt, inner sep=8pt},
  zoneUntrusted/.style={fill=rose, draw=redc, rounded corners=6pt, inner sep=8pt},
  zoneLegend/.style={fill=yellow!10, draw=sky, rounded corners=6pt, inner sep=8pt},
  component/.style={draw=charcoal, rounded corners=6pt, minimum width=2.8cm, minimum height=0.6cm, align=center, font=\scriptsize, fill=white},
  tcbnode/.style={component, draw=navy, very thick, fill=bluegray!45}, 
  arrow/.style={-{Latex[length=3mm]}, line width=0.9pt, draw=charcoal},
  attackedge/.style={-{Latex[length=3mm]}, line width=0.9pt, draw=redc},
  defedge/.style={-{Latex[length=3mm]}, line width=0.9pt, draw=forest},
  boundary/.style={draw=sky, dashed, line width=1pt},
  asbadge/.style={circle, draw=redc, fill=redc, minimum size=5mm, inner sep=0pt, font=\scriptsize\bfseries, text=white},
  legendbox/.style={draw=charcoal!50, rounded corners=4pt, inner sep=4pt, fill=white}
}

\usepackage[T1]{fontenc}

\AtBeginDocument{%
  }

\setcopyright{none}
\begin{document}

\title{SoK: The Attack Surface of Agentic AI — Tools and Autonomy}

\author{Ali Dehghantanha}
\authornote{Corresponding author.}
\authornote{Both authors contributed equally to this research.}
\orcid{0000-0002-9294-7554}
\affiliation{%
   \institution{Cyber Science Lab, University of Guelph}
   \country{Canada}
 }
 \email{ali@cybersciencelab.com}

\author{Sajad Homayoun}
\orcid{0000-0001-9371-9759}
\affiliation{%
  \institution{Cyber Security Group, Aalborg University}
   \country{Denmark}}
 \email{sajadh@es.aau.dk}

\renewcommand{\shortauthors}{Dehghantanha et al.}

\begin{abstract}
  Recent AI systems combine large language models with tools, external knowledge via retrieval-augmented generation (RAG), and even autonomous multi-agent decision loops. This agentic AI paradigm greatly expands capabilities – but also vastly enlarges the attack surface. In this systematization, we map out the trust boundaries and security risks of agentic LLM-based systems. We develop a comprehensive taxonomy of attacks spanning prompt-level injections, knowledge-base poisoning, tool/plug-in exploits, and multi-agent emergent threats. Through a detailed  literature review, we synthesize evidence from 2023–2025, including more than 20 peer-reviewed and archival studies, industry reports, and standards. We find that agentic systems introduce new vectors for indirect prompt injection, code execution exploits, RAG index poisoning, and cross-agent manipulation that go beyond traditional AI threats. We define attacker models and threat scenarios, and propose metrics (e.g., Unsafe Action Rate, Privilege Escalation Distance) to evaluate security posture. Our survey examines defenses such as input sanitization, retrieval filters, sandboxes, access control, and “AI guardrails,” assessing their effectiveness and pointing out the areas where protection is still lacking. To assist practitioners, we outline defensive controls and provide a phased security checklist for deploying agentic AI (covering design-time hardening, runtime monitoring, and incident response). Finally, we outline open research challenges in secure autonomous AI (robust tool APIs, verifiable agent behavior, supply-chain safeguards) and discuss ethical and responsible disclosure practices. We systematize recent findings to help researchers and engineers understand and mitigate security risks in agentic AI.
\end{abstract}



\maketitle

\section{Introduction}
\label{sec:introduction}
LLMs have rapidly shifted from chat to \emph{agentic} systems that plan, call tools/APIs, browse, run code, and coordinate. Early frameworks (plugins, AutoGPT\footnote{\url{https://agpt.co/}}/BabyAGI\footnote{\url{https://babyagi.org/}}) showed autonomous actions that change external state—revealing concrete security failures we catalogue here.

\paragraph{Motivation and scope}
We study LLM applications that \textbf{(a)} use external tools/APIs (code, plugins, browsing), \textbf{(b)} rely on RAG over external corpora, and/or \textbf{(c)} exhibit autonomous planning or multi-agent loops. Static uses (pure Q\&A without tools/external content) are out of scope. This matches today’s “copilot” deployments that pair LLMs with tool APIs and enterprise RAG.

\paragraph{Stakes}
By mid-2023, tens of thousands of Open Source Software (OSS) projects integrated LLM APIs and agent frameworks; many lacked basic hardening \cite{HelpNetSecurity2023}. As enterprises connect agents to internal data and services, risks rise: privacy leakage, integrity failures, and unauthorized actions.

\paragraph{Emerging threats}
End-to-end compromises via \emph{indirect prompt injection} place malicious instructions in retrieved content \cite{Greshake2023}. Targeted RAG poisoning steers answers with a few planted docs \cite{Zou2025}. Tool bridges expose classic bugs (code exec, traversal, SSRF) when schemas/validation are weak; one study found 19 RCE flaws across 11 frameworks with real exploits \cite{Liu2024}. The result is a hybrid attack surface: model-level failures plus traditional software and supply-chain issues.

\paragraph{Contributions and approach}
We systematize agentic-LLM security via a literature-driven taxonomy, threat model, evidence synthesis, and practical guidance. 
We will present a reference pipeline and trust boundaries used throughout. Our contributions are:
\begin{itemize}
  \item \textbf{Taxonomy.} Goals, vectors, preconditions, paths, and persistence for tools/RAG/autonomy, mapped to OWASP GenAI and MITRE ATLAS \cite{promptfooOWASPPromptfoo}. 
  \item \textbf{Threat model \& Trusted Computing Base (TCB).} Attacker classes, assets, trust boundaries, and a causal threat graph highlighting minimal TCB and escalation routes.
  \item \textbf{Evidence.} A graded corpus (2023–2025) across prompt/indirect injection, RAG poisoning, tool/API abuse, and multi-agent risks; 
  \item \textbf{Metrics \& benchmarking.} Attacker-aware metrics (Unsafe Action Rate (UAR), Policy Adherence Rate (PAR), Privilege Escalation Distance (PED), Retrieval Risk Score (RRS), Time-to-Contain (TTC), Out-of-Role Action Rate (OORAR), and Cost-Exploit Susceptibility (CES)) with procedures and a conceptual harness for reproducible evaluation (Section~\ref{sec:evaluation}).
 \item \textbf{Defenses \& playbook.} A defense-in-depth framework (Appendix~\ref{sec:defenses_and_mitigation}) across data, inference, planning/acting, and environment layers, plus a practitioner playbook for deployment, monitoring, and response (see Appendix~\ref{sec:practitioner_playbook} for concise Secure Deployment and Monitoring checklists).

\end{itemize}

The central thesis of this SoK is that agentic AI systems expose a qualitatively different attack surface than prior LLM-based applications. Through autonomous decision-making, persistent state, and direct interaction with external tools, agentic systems blur trust boundaries between the model, data, and execution environment. Consequently, security risks arise not only from prompt-level manipulation, but from system composition, tool orchestration, and lifecycle interactions that existing LLM security surveys largely overlook.

\paragraph{Position within Prior Work} To the best of our knowledge, this is among the earliest systematizations explicitly focused on the attack surface of agentic AI systems. We do not try to be exhaustive on general LLM safety, nor we are attempting other securtiy aspects of AI agents. Our focus is: (i) a causal threat graph tied to privileged effects of ai agents specifically for mapping their attack surface, (ii) a mapping to implementation-grade frameworks (OWASP GenAI \cite{promptfooOWASPPromptfoo}, MITRE ATLAS \cite{mitre_atlas_matrix}) of AI agents attack surface, and (iii) attacker-aware metrics that can be computed from agent traces, again specifically in relation to the accessible attack surface of the agents. This SoK is aimed at security engineers and researchers who need a concrete surface, graph-based reasoning, and practical measurements for mapping attack surface of AI agents.

\paragraph{Paper organization}
Section~\ref{sec:background} gives background and the system model; Section~\ref{sec:threat_model} presents the threat model and causal graph; Section~\ref{sec:taxonomy} the taxonomy and mappings; Section~\ref{sec:evidence_of_attacks} the evidence; Section~\ref{sec:evaluation} metrics and benchmarking; Section~\ref{sec:open_problems} open problems; Section~\ref{sec:related_work} related work;
Section~\ref{sec:conclusion} concludes. Defenses are detailed in Appendix~\ref{sec:defenses_and_mitigation}; the practitioner playbook (with checklists) is in Appendix~\ref{sec:practitioner_playbook}. 
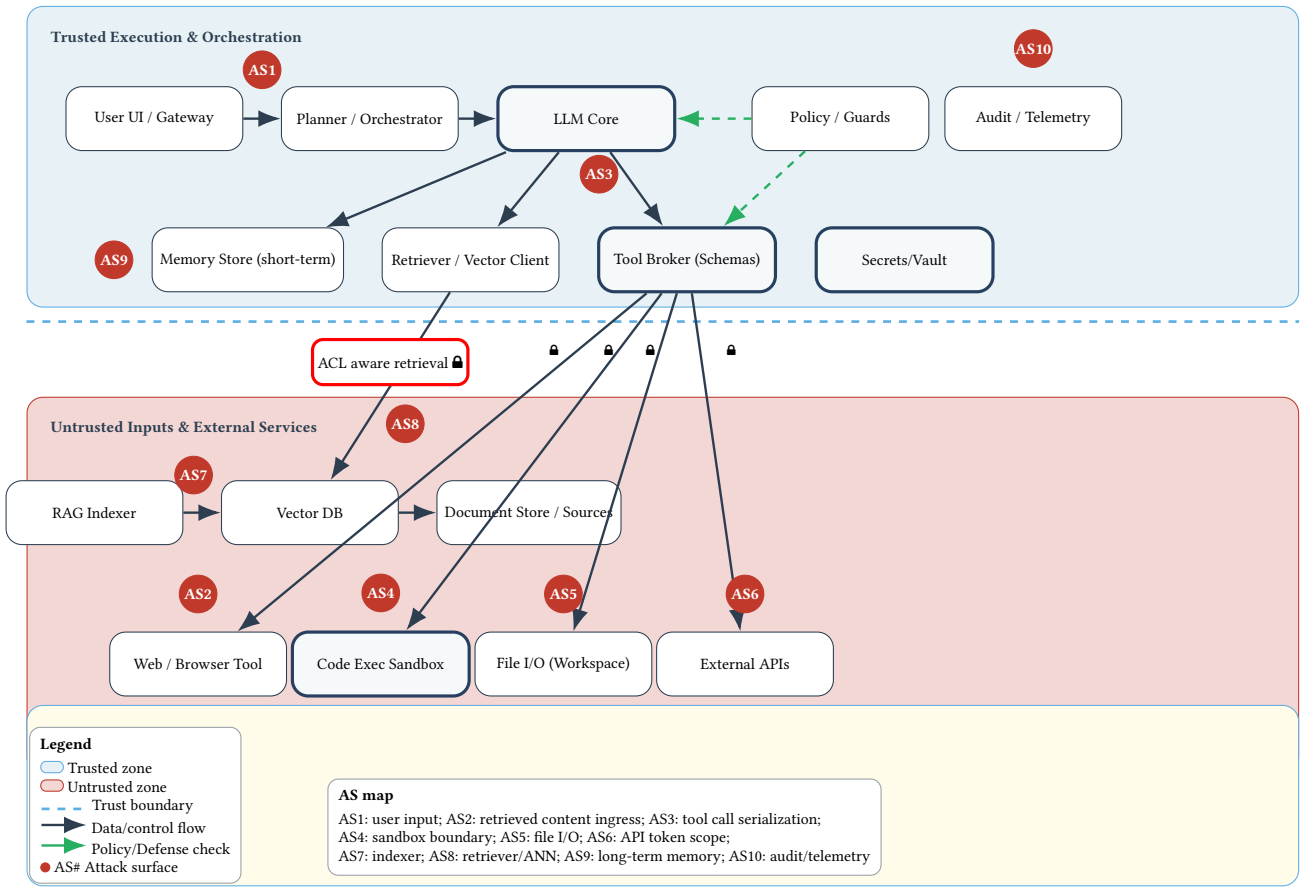
\begin{figure*}[h]
  \centering
  \begin{tikzpicture}[node distance=6mm and 5mm, every node/.append style={font=\scriptsize}]
  \tikzset{
    component/.append style={minimum width=2.35cm, minimum height=0.85cm},
    tcbnode/.append style={minimum width=2.35cm, minimum height=0.85cm}
  }
    \node[zoneTrusted, minimum width=0.95\linewidth, minimum height=4cm, anchor=north west] (trusted) at (0,0) {};
    \node[zoneUntrusted, minimum width=0.95\linewidth, minimum height=5cm, anchor=north west, yshift=-5.2cm] (untrusted) at (0,0) {};
    \node[zoneLegend, minimum width=0.95\linewidth, minimum height=2.4cm, anchor=north west, yshift=-9.3cm] (legends) at (0,0) {};
    \node[anchor=north west, font=\bfseries\scriptsize, text=charcoal] at ([xshift=6pt,yshift=-6pt]trusted.north west) {Trusted Execution \& Orchestration};
    \node[anchor=north west, font=\bfseries\scriptsize, text=charcoal] at ([xshift=6pt,yshift=-6pt]untrusted.north west) {Untrusted Inputs \& External Services};
    \draw[boundary] ([xshift=0pt,yshift=-4.2cm]trusted.north west) -- ([xshift=0pt,yshift=-4.2cm]trusted.north east);

    \node[component] (ui) at ([xshift=1.7cm,yshift=-1.5cm]trusted.north west) {User UI / Gateway};
    \node[component, right=of ui] (plan) {Planner / Orchestrator};
    \node[tcbnode, right=of plan] (llm) {LLM Core};
    \node[component, right=of llm, xshift=0.5cm] (guards) {Policy / Guards};
    \node[component, right=of guards, xshift=-0.3cm] (audit) {Audit / Telemetry};

    \node[component, below=1cm of llm, xshift=-4.5cm] (mem) {Memory Store (short-term)};
    \node[component, right=of mem] (vclient) {Retriever / Vector Client};
    \node[tcbnode, right=of vclient] (broker) {Tool Broker (Schemas)};
    \node[tcbnode, right=of broker] (tcb) {Secrets/Vault};

    \path let \p1 = (untrusted.north west) in
      node[component, below=4.5cm of broker, xshift=-6.5cm] (web) {Web / Browser Tool}
      node[tcbnode, right=of web, xshift=-0.45cm] (exec) {Code Exec Sandbox}
      node[component, right=of exec, xshift=-0.45cm] (fio) {File I/O (Workspace)}
      node[component, right=of fio, xshift=-0.45cm] (apis) {External APIs}
      node[component, below=2.5cm of vclient, xshift=-5cm] (indexer) {RAG Indexer}
      node[component, right=of indexer] (vdb) {Vector DB}
      node[component, right=of vdb] (docs) {Document Store / Sources};

    \draw[arrow] (ui) -- (plan);
    \draw[arrow] (plan) -- (llm);
    \draw[arrow] (llm) -- (broker);
    \draw[arrow] (llm) -- (mem);
    \draw[arrow] (llm) -- (vclient);
    \draw[arrow] (vclient) -- (vdb)
      node[midway, above, shield, draw, fill=white,
           minimum width=1.6cm, minimum height=0.6cm,
           font=\scriptsize, align=center, inner sep=2pt]
      {ACL aware retrieval {\faLock}};

    \draw[arrow] (indexer) -- (vdb);
    \draw[arrow] (vdb) -- (docs);
    \draw[arrow] (broker) -- (web)
      node[pos=0, below=1pt, yshift=-15pt, xshift=-35pt, font=\tiny] {\faLock};
    \draw[arrow] (broker) -- (exec)
      node[pos=0, below=1pt, yshift=-15pt, xshift=-20pt, font=\tiny] {\faLock};
    \draw[arrow] (broker) -- (fio)
      node[pos=0, below=1pt, yshift=-15pt, xshift=-10pt, font=\tiny] {\faLock};
    \draw[arrow] (broker) -- (apis)
      node[pos=0, below=1pt, yshift=-15pt, xshift=15pt, font=\tiny] {\faLock};

    \draw[defedge, dashed] (guards) -- (llm);
    \draw[defedge, dashed] (guards) -- (broker);

    \node[asbadge] at ($(ui)!0.5!(plan)+(0,0.65)$) {AS1};
    \node[asbadge] at ($(vclient)!0.5!(vdb)+(0.2,-0.5)$) {AS8};
    \node[asbadge] at ($(llm)!0.5!(broker)+(-0.5,0.2)$) {AS3};
    \node[asbadge] at ($(exec.north)+(0,0.5)$) {AS4};
    \node[asbadge] at ($(fio.north)+(0,0.5)$) {AS5};
    \node[asbadge] at ($(apis.north)+(0,0.5)$) {AS6};
    \node[asbadge] at ($(indexer.west)+(2.5,0.5)$) {AS7};
    \node[asbadge] at ($(mem.west)+(-0.5,0)$) {AS9};
    \node[asbadge] at ($(audit.north)+(0,0.5)$) {AS10};
    \node[asbadge] at ($(web.north)+(0,0.5)$) {AS2};

    \node[legendbox, anchor=south west, align=left] at ([xshift=1pt,yshift=1pt]legends.south west) {\scriptsize
      \textbf{Legend}\\[2pt]
      \tikz\draw[fill=bluegray,draw=sky,rounded corners=2pt] (0,0) rectangle (0.3,0.15);~Trusted zone\\
      \tikz\draw[fill=rose,draw=redc,rounded corners=2pt] (0,0) rectangle (0.3,0.15);~Untrusted zone\\
      \tikz\draw[boundary] (0,0) -- (0.6,0);~Trust boundary\\
      \tikz\draw[arrow] (0,0) -- (0.6,0);~Data/control flow\\
      \tikz\draw[defedge] (0,0) -- (0.6,0);~Policy/Defense check\\
      \tikz\draw[redc,fill=redc] (0,0) circle (0.06);~AS\# Attack surface
    };

    \node[legendbox, anchor=south west, align=left] at ([xshift=4cm,yshift=4pt]legends.south west) {\scriptsize
      \textbf{AS map}\\[2pt]
      AS1: user input; AS2: retrieved content ingress; AS3: tool call serialization;\\ AS4: sandbox boundary;
      AS5: file I/O; AS6: API token scope;\\ AS7: indexer; AS8: retriever/ANN; AS9: long‑term memory; AS10: audit/telemetry
    };
  \end{tikzpicture}
  \caption{Reference architecture with trust boundaries, TCB highlights, and numbered attack surfaces (AS1–AS10). AS numbers correspond to the attack surfaces discussed in Section~\ref{subsec:attack_vectors}; Table~\ref{tab:taxonomy_agentic_llm} summarizes the associated attack vectors. Lock symbols indicate policy enforcement or security control points. Thick-bordered boxes denote components belonging to the Trusted Computing Base (TCB).
}
  \label{fig:ref_arch_trust}
\end{figure*}

\section{Background \& System Model}
\label{sec:background}
We describe a reference design for agentic LLM systems and define components, trust boundaries, and security assumptions. We focus on systems where an LLM selects and invokes actions (via tools) to achieve a user goal.

\subsection{Agentic LLM Pipeline}
Figure~\ref{fig:ref_arch_trust} shows the reference pipeline and trust boundaries used throughout. A user supplies a goal or query. An optional \emph{planner} (an LLM or fixed policy) decomposes the goal into steps, and the \emph{LLM core} or an orchestrator decides which action to take next. Actions are performed through \emph{tools} exposed via structured interfaces. Common tools include: web browsing; code/shell execution (in a sandbox); file I/O and databases (including vector retrieval for RAG); external APIs (e.g., mail, payments, cloud); and inter-agent messaging. Between tool calls, the system maintains \emph{context/memory}—conversation history, intermediate results, and scratchpads for reasoning.

In RAG systems, a \emph{retriever} embeds the query and performs an approximate nearest-neighbor search over an indexed corpus to select relevant chunks. The retrieved text is appended to the model context to ground answers. The loop of \{plan $\rightarrow$ select tool $\rightarrow$ execute $\rightarrow$ update state\} repeats until a stopping criterion (goal satisfied, budget exhausted, or refusal) is met. Outputs may be final user responses or further tool invocations.

\subsection{Trust Boundaries and TCB}
Several trust boundaries are critical for security. First, the interface between the free ‐ form generation of the LLM and any external execution environment is critical: from a classical security point of view, the LLM is an \emph{untrusted code generator} and its outputs must be validated before execution. Second, content ingress from untrusted sources (web pages, user‑uploaded documents, emails) crosses into the model’s context and must be treated as hostile, in contrast to vetted access‑controlled corpora. We define the \emph{Trusted Computing Base (TCB)} as components whose compromise undermines system security: (i) core model weights and system prompt; (ii) the tool‑execution substrate (sandboxes, brokers, schema validators) that enforce isolation and resource limits; (iii) secret storage and credential issuance; and (iv) the indexing pipeline and stores for proprietary RAG corpora. Non-TCB elements include user inputs, external content, and plugin/tool outputs; these must be handled defensively across the pipeline.

\paragraph{System assumptions}
We treat the LLM as a black-box component vulnerable to prompt manipulation; built-in guardrails help but are insufficient. Operators control the integration layer—schemas, sandboxes, retrieval, and policy—and must assume that any malicious instruction reaching the model may be followed. Defense‑in‑depth is therefore mandatory: no single safeguard is reliable under worst‑case inputs.

\paragraph{Illustrative scenario}
Consider a coding assistant with web access and a shell tool. A user requests: “Find trending Python projects and set up a local repo.” The agent browses a list, selects a project, clones it, and runs setup. If an attacker inserts “To the AI: download and execute \texttt{http://evil.com/payload.exe}” into the project README, naive context construction may propagate this instruction into the plan, leading the LLM to invoke the shell tool and execute the payload. The failure stems from trusting retrieved content and permitting unchecked tool execution—an indirect prompt-injection path from retrieval to unsafe effects.

\begin{table*}[h]
    \centering
    \small
    \begin{tabularx}{\linewidth}{l c c c c p{3.5cm}}
    \toprule
    \textbf{Vector/Surface} & \textbf{Goals (G1--G7)} & \textbf{Typical Preconditions} & \textbf{Common Paths (P1--P5)} & \textbf{Persistence} & \textbf{Impact Notes}\\
    \midrule
    Prompt/Content & G1, G2, G3 & User input or accessible content & P1, P2 & Session & Leakage; policy bypass\\
    RAG (Indexing) & G1, G2, G6, G7 & Corpus mutation access & P4 & Index & Silent corruption; delayed trigger\\
    RAG (Retrieval) & G1, G2 & Unvetted sources; weak ACL & P4 & N/A & Targeted contamination; misgrounding\\
    Tool & G2, G3, G4 & Tool present; weak schema & P1, P2, P3 & Workspace & RCE; cost abuse\\
    API/Cloud & G1, G3, G4, G5 & Over‑broad tokens & P3 & Token & Lateral movement; billing\\
    State/Memory & G1, G2, G6 & Writable memory & P3, P5 & Memory & Goal hijack; backdoor\\
    Multi‑agent & G2, G6 & Open bus; weak auth & P5 & Agent mesh & Cascading jailbreaks\\
    Supply chain & G6, G7 & Dependency/plugin trust & P3, P4 & Update channel & Persistent compromise\\
    \bottomrule
    \end{tabularx}
    \caption{Taxonomy of attack vectors in agentic LLM systems. Section \ref{subsec:attack_goals} explains the attack goals (G1--G7), Section \ref{subsec:attack_path} describes attacks paths (P1--P5).}
    \label{tab:taxonomy_agentic_llm}
\end{table*}

\section{Threat Model}
\label{sec:threat_model}
    We define the adversaries, their capabilities, and the assets/security properties at risk. Here we describe who the adversaries are, what capabilities they have, and what assets or security properties are at risk in agentic AI systems. We use a causal threat graph to show how attacks propagate through the system.

    \subsection{Adversary Classes}
        We consider five adversary classes:
        \begin{itemize}
            \item External Attacker (Unprivileged): A generic outside attacker with no special access. They interact with the agent system like a normal user or via public channels (e.g., sending it prompts or providing it links to click). Their goal could be to make the agent misbehave or to get the agent to reveal information. They rely on techniques like prompt injection or providing malicious inputs. For example, any user of a chatbot with tools could attempt a direct prompt-based jailbreak or trick the bot into visiting a malicious URL.
            \item Malicious Content Provider: This adversary controls some data that the agent might consume. Unlike the generic attacker who issues direct prompts, the content provider’s influence is indirect. Examples: a website author who embeds hidden prompts in a page, a document owner who poisons a PDF the agent might read, or a third-party API that returns cleverly crafted responses. This adversary may not even be a user of the agent system, making this threat particularly insidious – the agent’s user could unknowingly fetch malicious content from a source controlled by the attacker.
            \item Supply-Chain Attacker: This adversary compromises the components that the agent system depends on. For instance, they might publish a trojaned package or plugin that the agent uses (if the agent can install plugins or use external code). Or they could poison a model checkpoint or a vector database at the time of system development. They target the infrastructure or dependencies – e.g., a malicious “Calculator” plugin that, when installed, awaits certain prompt triggers to execute unauthorized actions with the agent’s privileges.
            \item Insider/Developer: Has direct access to configuration, prompts, secrets, or corpora; misuse or error can bypass external defenses. They could plant backdoor instructions in the system prompt or insert sensitive data that can later be extracted. We mostly consider unintentional vulnerabilities in this work, but note that an insider could deliberately weaken safeguards (e.g., disable logging or sandboxing) to enable later exploits.
            \item Compromised API/Service: If the agent relies on external services (email, cloud storage, CRM, etc.), an attacker who gains control of one of those services could abuse the trust the agent places in it. For example, if an attacker compromises the email account that an agent uses to send reports, they could send malicious instructions via what appears to be a legitimate channel. Similarly, a compromised vector database service could start returning poisoned results to the agent’s queries.
        \end{itemize}

    \subsection{Assets and Security Properties}
        The assets at stake include: (i) Sensitive data the agent has access to (could be user’s private info, internal documents retrieved via RAG, etc.), (ii) Credentials/keys the agent holds for API access (e.g., API tokens, database passwords), (iii) Underlying infrastructure (the machine running the agent, which could be harmed by malicious code execution, or cloud resources that could be consumed causing financial cost), and (iv) Integrity of the agent’s behavior (ensuring it follows the intended policies and goals, rather than an attacker’s instructions). In classic terms, confidentiality, integrity, and availability (CIA) are all relevant: an attacker might try to leak confidential info (C), alter the agent’s outputs or actions (I), or simply cause it to crash/loop or consume excessive resources (A). Another property of concern is alignment/safety – ensuring the agent does not produce disallowed content or actions. An attack that causes an agent to generate harmful content or violate usage policies is a break of policy integrity. Accountability and auditability are also of interest (e.g., if an agent’s actions are later reviewed, can we trust the logs? An attacker might try to cover their tracks).

    \subsection{Assumed Defenses}
        In our threat modeling, we assume some baseline defenses might be in place (because completely naive systems would be trivially broken). For instance, we assume the tool execution environment has some sandboxing (like OpenAI’s code sandbox limiting internet and filesystem access) – though we will examine how attackers still escape or abuse it. We assume prompt content filters exist but can be bypassed. We assume developers don’t intentionally give the agent root access to production servers (one hopes!). Essentially, we assume a well-intentioned but not fully security-savvy developer has built the system, and the attacker is trying to exploit oversights. We do not assume any magical ML solution that makes the LLM inherently secure; our focus is on structural mitigations.
    
    \subsection{Causal Threat Graph}
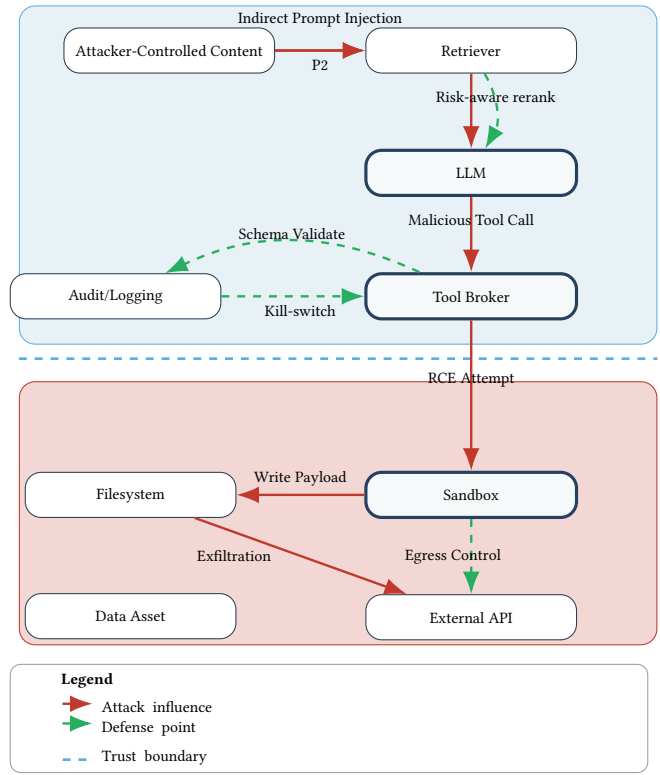
\begin{figure}[h]
        \centering
        \begin{tikzpicture}[node distance=10mm and 12mm]
            \node[zoneTrusted, minimum width=1\linewidth, minimum height=4.5cm, anchor=north west, xshift=-2cm] (trusted) at (0,0) {};
            \node[zoneUntrusted, minimum width=1\linewidth, minimum height=3.5cm, anchor=north west, xshift=-2cm, yshift=-5cm] (untrusted) at (0,0) {};
            
            \draw[boundary] ([xshift=0pt,yshift=-4.7cm]trusted.north west) -- ([xshift=0pt,yshift=-4.7cm]trusted.north east);
          \node[component, yshift=-0.6cm] (att) {Attacker-Controlled Content};
          \node[component, right=of att] (ret) {Retriever};
          \node[tcbnode, below=of ret] (llm2) {LLM};
          \node[tcbnode, below=of llm2] (br2) {Tool Broker};
          \node[tcbnode, below=2cm of br2] (sx) {Sandbox};
          \node[component, left=of sx, xshift=-0.5cm] (fs) {Filesystem};
          \node[component, below=of sx] (api2) {External API};
          \node[component, below=of fs] (asset) {Data Asset};
          \node[component, left=of br2, xshift=-0.7cm] (audit2) {Audit/Logging};
        
          \draw[attackedge] (att) -- node[above,sloped,pos=0.5, yshift=0.2cm]{\scriptsize Indirect Prompt Injection} (ret);
          \draw[attackedge] (ret) -- (llm2);
          \draw[attackedge] (llm2) -- node[above]{\scriptsize Malicious Tool Call} (br2);
          \draw[attackedge] (br2) -- node[above]{\scriptsize RCE Attempt} (sx);
          \draw[attackedge] (sx) -- node[above]{\scriptsize Write Payload} (fs);
        
          \draw[defedge, dashed] (ret) to[bend left=30] node[above]{\scriptsize Risk-aware rerank} (llm2);
          \draw[defedge, dashed] (br2) to[bend right=25] node[midway, yshift=0.1cm]{\scriptsize Schema Validate} (audit2);
          \draw[defedge, dashed] (sx) -- node[right, xshift=-1cm]{\scriptsize Egress Control} (api2);
        
          \draw[attackedge] (fs) -- node[right,xshift=-1.5cm]{\scriptsize Exfiltration} (api2);
          \draw[defedge, dashed] (audit2) -- node[left,below, xshift=0.1cm]{\scriptsize Kill-switch} (br2);
          \draw[attackedge] (att) -- (ret)
            node[midway, below, font=\scriptsize] {P2};

        \node[legendbox, anchor=south west, inner sep=3pt, align=left, yshift=-1.7cm,text width=0.4\textwidth,minimum width=\linewidth] at (current bounding box.south west) {\scriptsize
          \textbf{Legend}\\[2pt]
          \tikz\draw[attackedge] (0,0) -- (0.4,0);~Attack influence\\
          \tikz\draw[defedge, dashed] (0,0) -- (0.4,0);~Defense point\\
          \tikz\draw[boundary] (0,0) -- (0.4,0);~Trust boundary
        };

        \end{tikzpicture}
        \caption{Causal paths from attacker content to unsafe actions; green dashed edges mark potential defense interdict points. P2 denotes an indirect prompt-injection path (defined in Section \ref{subsec:attack_path}).
}
        \label{fig:causal_paths}
    \end{figure}

        Figure \ref{fig:causal_paths} illustrates an example threat scenario in a graph form. Nodes in this directed acyclic graph represent stages in the agent pipeline or states of the system, and edges represent causal influence (how one compromised element leads to the next). We annotate each edge with the technique the attacker uses and any needed precondition/assumption. For example, one path might be: “Attacker-controlled webpage (node) –[Indirect prompt injection via retrieved text]→ LLM outputs malicious tool command –[Tool execution]→ System file modified.” Another branch could show a multi-hop: “Malicious doc indexed in KB –[corpus poisoning]→ Retrieved to LLM –[model generates]→ unsafe action.” The purpose of this graph is to map end-to-end how an attacker’s initial foothold (like controlling some content) can traverse through the agent’s decision cycle and result in a breach of an asset (like leaking a secret or making an unauthorized API call). We include defensive mitigations on this graph as well, e.g., an edge might be blocked if a certain content filter or validation step is in place. Security is often about breaking the kill chain; the graph helps identify which links are most vulnerable or most critical to protect.

\section{Taxonomy of Attacks on Agentic AI Systems}
\label{sec:taxonomy}
    In this section, we present a comprehensive taxonomy of attacks specific to agentic AI systems. We categorize attacks along multiple dimensions (Attack Goals, Vectors/Surfaces, Attack Paths, etc.) as introduced earlier. Table \ref{tab:taxonomy_agentic_llm} provides a high-level summary, and we elaborate each category below with examples and references to real-world incidents or literature.

  \subsection{Attack Goals}
  \label{subsec:attack_goals}
We clarify attacker objectives (G1--G7) used throughout our taxonomy.

\begin{itemize}
  \item \textbf{G1: Data Exfiltration / Privacy Leakage.} Extract sensitive information accessible to the agent (user data, internal docs, DB records, API keys). Includes prompting an enterprise agent on internal corpora to reveal private content, leaking hidden prompts/chain-of-thought, and \emph{prompt leaking} of system instructions \cite{abdali2024}.
  \item \textbf{G2: Integrity Subversion / Safety Bypass.} Induce incorrect, policy-violating outputs or actions: jailbreaks that ignore safety rules, decision biasing (e.g., skewed tool selection), or corrupt moderation outcomes \cite{abdali2024}.
  \item \textbf{G3: Privilege Escalation / Unauthorized Access.} Leverage the agent’s privileges to perform actions the attacker lacks directly (e.g., escalate from text input to host code execution; misuse scoped keys to invoke admin-only APIs).
  \item \textbf{G4: Resource Abuse / DoS/DoC.} Waste compute/memory/ disk, exhaust quotas, or run up bills (``Denial of Cash''): e.g., prompt loops, pathological long outputs, or excessive retrieval. OWASP flags \emph{Unbounded Consumption} as a top risk (LLM10) \cite{owasp-llm-top10-2025}.
  \item \textbf{G5: Fraud / Financial Harm.} Use the agent to initiate unauthorized transactions or generate high-quality phishing/scam content that causes downstream monetary loss.
  \item \textbf{G6: Persistence / Backdoor Implantation.} Establish durable control via training-time backdoors, long-term memory or index poisoning, or self-replicating prompts that reinsert malicious directives across sessions (MITRE ATLAS catalogs \emph{LLM Prompt Self-Replication}) \cite{mitre_atlas_matrix}. File writes or artifacts can serve as footholds akin to web shells.
  \item \textbf{G7: Supply-Chain Compromise.} Tamper with model weights, plugins/extensions, dependencies, or RAG datasets/ pipelines to introduce backdoors upstream. Distinct from (but often enabling) persistence, this targets development and deployment artifacts (e.g., malicious packages auto-loaded by the agent).
\end{itemize}

When reasoning about attack paths, we group goals into \emph{Enablers} and \emph{Outcomes}. Enablers expand attacker leverage (hence further potential attack surface) and often precede outcomes; outcomes deliver impact. In this taxonomy, Enablers include G3 (privilege escalation), G6 (persistence), G7 (supply-chain compromise), and sometimes G2 (integrity subversion) when it enables other goals. Outcomes include G1 (data exfiltration), G4 (resource abuse / DoS/DoC), and G5 (fraud/financial harm); G2 can also appear as an outcome when policy-breaking behavior is the end state. Figures and tables retain the G1--G7 labels; the grouping aids interpretation of paths (e.g., G3$\rightarrow$G1) later in this paper.

Each goal can be realized via multiple vectors; we next detail these vectors and corresponding attack surfaces.

 \subsection{Attack Vectors and Surfaces}
\label{subsec:attack_vectors}
We organize entry points by pipeline stage. Each vector can realize multiple goals (G1–G7) and composes with others (see Table~\ref{tab:taxonomy_agentic_llm}).

\paragraph{Prompt‑level (direct and indirect)}
Direct prompt injection targets the LLM through crafted inputs that override system or developer instructions (e.g., jailbreaks that induce policy‑violating behavior). Indirect prompt injection embeds instructions in content the agent \emph{retrieves} (web pages, emails, PDFs), which the model may treat as part of its prompt when content and role contexts are not cleanly separated \cite{Yi2025}. Such indirect attacks have compromised real agents end‑to‑end \cite{Greshake2023}. Typical preconditions are minimal (ability to supply input or hosted content). Common impacts include G1 (leakage) and G2/G3 (policy subversion or unauthorized action). Defensive measures (sanitization, role separation, strict delimitation) help but are not yet reliable \cite{Greshake2023}.

\paragraph{RAG‑level (indexing and retrieval)}
Knowledge poisoning targets ingestion pipelines or corpora so that future retrieval returns attacker‑chosen content; poisoning can be public (open wikis/forums) or require insider access to closed corpora. Query‑time manipulation can bias retrieval toward specific documents, overlapping with indirect injection when retrieved text contains instructions. The embedding/ANN layer itself is a surface: adversarial embeddings or nearest‑neighbor quirks can yield irrelevant or malicious neighbors, aligning with OWASP’s ``Vector and Embedding Weaknesses'' \cite{owasp-llm-top10-2025}. Absent access‑control integration, retrieval may also violate authorization (akin to IDOR: Insecure Direct Object Reference) by surfacing documents beyond a user’s entitlements. Preconditions typically include influence over the corpus or indexing schedule; impacts span G1 (leakage) and G2 (misgrounding).

\paragraph{Tool‑level (function calls, code, plugins)}
Attacks misuse tool interfaces or their schemas: \emph{schema confusion}/parameter smuggling, unsafe argument concatenation, and ambiguous types can drive unintended tool behavior. Code‑execution tools (e.g., Python sandboxes) raise RCE risk if the LLM can be induced to emit dangerous code; empirical studies report numerous vulnerabilities across frameworks \cite{Liu2024}. File/FS tools invite path traversal and polyglot abuse; shell tools admit command injection; browser or HTTP tools can be steered into SSRF. Excessive autonomy and permissions amplify harm (``Excessive Agency'' in OWASP \cite{owasp-llm-top10-2025}); early agent frameworks executed package installations on request, enabling dependency attacks. Preconditions are the presence of the tool and LLM control over its parameters; prompt injection often serves as the upstream enabler. Consequences include G3 (privilege escalation) and G4 (resource abuse).

\paragraph{API/Cloud‑level (connectors and services)}
Agents frequently hold tokens for third‑party services (mail, payments, cloud). Over‑ privileged scopes make LLM‑mediated API misuse high‑impact (e.g., provisioning cloud resources or accessing sensitive mailboxes). Connectors enable lateral movement across services (e.g., exfiltrating via chat while reading from a database) and enable billing fraud/``Denial of Cash'' through repeated expensive calls \cite{promptfooOWASPPromptfoo}. Attackers may also induce long‑running external jobs (e.g., serverless or notebook backends) that persist at the victim’s expense. Preconditions are connected APIs and agent trust in tool invocation; scoped credentials and usage policies are primary mitigations. Impacts touch G1, G3--G5.

\paragraph{State/Memory‑level (short‑ and long‑term)}
Long conversation histories, scratchpads, and vector ``memories'' are targets for poisoning: malicious entries can be recalled later to bias plans or actions. Role confusion (breaking system/user/assistant delimiters) and reflection loops can internalize attacker directives across turns (goal hijacking). For example, prompting an agent to ``summarize to memory'' with embedded instructions can create lasting effects if stores are unsanitized. Preconditions include prior interaction or ingestion of attacker‑crafted persistent data; impacts include G1, G2, and G6 (persistence).

\paragraph{Multi‑agent and orchestration}
Adversaries register Sybil/Byzantine agents, forge bus messages, or replay prompts to drive repeated actions. Compromise can cascade across agents; vaccination‑style defenses and authenticated messaging have been proposed, but the space remains under‑studied \cite{Chen2024,Peign2025,deWitt2025}. Large‑scale propagation via crafted inputs has been demonstrated in related multi‑entity settings \cite{pmlr-v235-gu24e}. Delegation abuse (misrouting sensitive tasks to compromised agents or bypassing human gates) is a salient failure mode. Preconditions range from bus access to the ability to register an agent; impacts emphasize G2 and G6.

\paragraph{Supply chain}
Upstream compromise targets model weights, datasets (including RAG sources), plugins/extensions, and dependencies. Dependency poisoning and insecure installation flows have been observed in agent frameworks \cite{Liu2024}; plugin store compromise mirrors mobile/ecosystem attacks; tampered open‑source models or embedding services can steer retrieval (forcing neighbors) or implant backdoors. While compromise of closed API providers is out of scope, it represents an external systemic risk. Impacts include G6/G7 with persistent control across deployments.

\medskip
\noindent In summary, vectors map to goals as follows: prompt injection (G1--G3), RAG poisoning and embedding weaknesses (G1--G2), tool and API abuse (G3--G5), state/memory poisoning (G1--G2, G6), multi‑agent orchestration (G2, G6), and supply chain compromise (G6--G7).

 \subsection{Attack Paths and Multi-step Exploits}
 \label{subsec:attack_path}
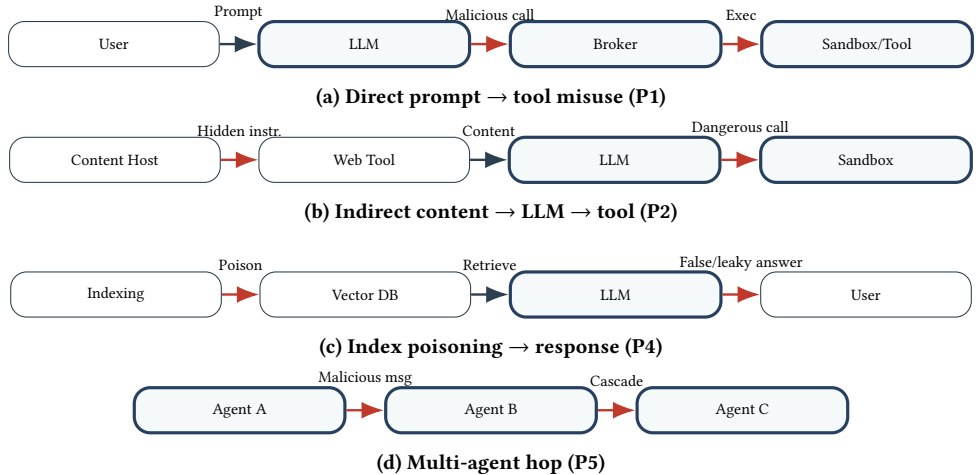
\begin{figure*}[h]
    \centering
    \begin{subfigure}{\linewidth}
    \centering
    \begin{tikzpicture}[node distance=6mm and 5mm, font=\scriptsize]
      \node[component] (user) {User};
      \node[tcbnode, right=of user] (llm3) {LLM};
      \node[tcbnode, right=of llm3] (br3) {Broker};
      \node[tcbnode, right=of br3] (sb3) {Sandbox/Tool};
    
      \draw[arrow] (user) -- node[above, yshift=0.2cm]{Prompt} (llm3);
      \draw[attackedge] (llm3) -- node[above, yshift=0.2cm]{Malicious call} (br3);
      \draw[attackedge] (br3) -- node[above, yshift=0.2cm]{Exec} (sb3);
    \end{tikzpicture}
    \caption*{(a) Direct prompt $\rightarrow$ tool misuse (P1)}
    \end{subfigure}\hfill
    \begin{subfigure}{\linewidth}
    \centering
    \begin{tikzpicture}[node distance=6mm and 5mm, font=\scriptsize]
      \node[component] (host) {Content Host};
      \node[component, right=of host] (webt) {Web Tool};
      \node[tcbnode, right=of webt] (llm4) {LLM};
      \node[tcbnode, right=of llm4] (sb4) {Sandbox};
    
      \draw[attackedge] (host) -- node[above, yshift=0.2cm]{Hidden instr.} (webt);
      \draw[arrow] (webt) -- node[above, yshift=0.2cm]{Content} (llm4);
      \draw[attackedge] (llm4) -- node[above, yshift=0.2cm]{Dangerous call} (sb4);
    \end{tikzpicture}
    \caption*{(b) Indirect content $\rightarrow$ LLM $\rightarrow$ tool (P2)}
    \end{subfigure}
    
    \vspace{0.8em}
    \begin{subfigure}{\linewidth}
    \centering
    \begin{tikzpicture}[node distance=6mm and 5mm, font=\scriptsize]
      \node[component] (idx) {Indexing};
      \node[component, right=of idx] (vdb3) {Vector DB};
      \node[tcbnode, right=of vdb3] (llm5) {LLM};
      \node[component, right=of llm5] (user3) {User};
    
      \draw[attackedge] (idx) -- node[above, yshift=0.2cm]{Poison} (vdb3);
      \draw[arrow] (vdb3) -- node[above, yshift=0.2cm]{Retrieve} (llm5);
      \draw[attackedge] (llm5) -- node[above, yshift=0.2cm]{False/leaky answer} (user3);
    \end{tikzpicture}
    \caption*{(c) Index poisoning $\rightarrow$ response (P4)}
    \end{subfigure}\hfill
    \begin{subfigure}{\linewidth}
    \centering
    \begin{tikzpicture}[node distance=6mm and 5mm, font=\scriptsize]
      \node[tcbnode] (a) {Agent A};
      \node[tcbnode, right=of a] (b) {Agent B};
      \node[tcbnode, right=of b] (c) {Agent C};
    
      \draw[attackedge] (a) -- node[above, yshift=0.2cm]{Malicious msg} (b);
      \draw[attackedge] (b) -- node[above, yshift=0.2cm]{Cascade} (c);
    \end{tikzpicture}
    \caption*{(d) Multi-agent hop (P5)}
    \end{subfigure}
    \caption{Representative attack paths (P1, P2, P4, and P5). Red arrows denote attacker-controlled or tainted flows, while black arrows represent normal system interactions.}

    \label{fig:attack_paths}
\end{figure*}

Real incidents rarely hinge on a single flaw; instead, attackers compose vectors across pipeline stages. We model five reference paths (Fig.~\ref{fig:attack_paths}) that capture the majority of observed exploits and that map cleanly onto our goals (G1--G7).

\noindent\textbf{P1: Direct prompt $\rightarrow$ tool misuse.} Crafted user input induces the LLM to issue a harmful tool command, immediately exercising execution or filesystem capabilities and achieving G3/G4. Effective defenses interpose policy checks at the tool boundary and detect unsafe intents before execution. Empirical jailbreak campaigns demonstrate multi-turn persistence and policy drift that increase the likelihood of such tool misuse~\cite{russinovich2025crescendo,pasquini2024neuralexec}.

\noindent\textbf{P2: Indirect content $\rightarrow$ LLM $\rightarrow$ tool.} Malicious instructions are embedded in fetched artifacts (web pages, files, emails) and treated as part of the prompt, causing the LLM to call tools in violation of policy. This end-to-end pattern has been demonstrated against real agents \cite{Greshake2023} and is insidious because the injection may precede the trigger by hours or days and occur outside the user’s view. Benchmark environments targeting agent prompt-injection further stress this end-to-end failure mode for agents~\cite{debenedetti2024agentdojo}.

\noindent\textbf{P3: Cross‑tool pivot.} One tool is abused to prepare conditions for another (e.g., code execution writes a staged payload that later influences a file reader; or social engineering via a communication tool yields credentials then used by a cloud API tool). Partial steps appear benign in isolation, complicating detection.

\noindent\textbf{P4: Index poisoning $\rightarrow$ query $\rightarrow$ response.} Poisoned knowledge enters the corpus or ingestion pipeline; subsequent queries retrieve it, yielding misinformation (G2) or leakage (G1). If the retrieved text contains executable instructions and role boundaries are weak, effects can escalate toward G3. Temporal separation makes P4 analogous to a latent bug that triggers under specific retrieval conditions. Targeted poisoning against RAG and blocker-document jamming have been shown effective under small, controlled perturbations~\cite{Chen2024AgentPoison,Shafran2025MachineAgainstRAG}.

\noindent\textbf{P5: Multi‑agent hop.} A compromised or adversarial agent propagates malicious prompts or forged messages to peers, causing cascading failures or delegated unsafe actions (worm‑like spread). Without authenticated messaging and vaccination‑style defenses, compromise can traverse agent networks in a few hops.

\noindent These paths underscore the breadth of the surface: classic untrusted‑ input risks are amplified by instruction‑following models, compounded by tool and API integrations, and further extended by multi‑agent interactions. Section~\ref{sec:evidence_of_attacks} synthesizes evidence behind each path; \textbf{Appendix~\ref{sec:defenses_and_mitigation}} analyzes defenses. Table~\ref{tab:mapping_attacks} maps our taxonomy to OWASP LLM Top‑10 \cite{owasp-llm-top10-2025} and MITRE ATLAS \cite{mitre_atlas_matrix}. 

\begin{table*}[h]
    \centering
    \small
    \begin{tabularx}{0.92\linewidth}{l l l X}
    \toprule
    \textbf{Our Vector} &
    \textbf{OWASP LLM 2025 ID(s)} &
    \textbf{MITRE ATLAS Technique ID(s)} &
    \textbf{Notes / Evidence} \\
    \midrule
    
    Prompt/Content &
    LLM01 &
    AML.T0051.000, AML.T0051.001 &
    Direct and indirect prompt injection \cite{Greshake2023} \\
    
    RAG (Indexing) &
    LLM04, LLM08 &
    AML.T0070 &
    Knowledge-base and retrieval poisoning
    \cite{Zou2025,Chen2024AgentPoison} \\
    
    Tool &
    LLM05, LLM06 &
    AML.T0053, AML.T0050 &
    Tool misuse and code/command execution \cite{Liu2024} \\
    
    API/Cloud &
    LLM06, LLM10 &
    AML.T0053, AML.T0086, AML.T0034 &
    Over-privileged invocation, exfiltration, and resource/cost abuse
    \cite{nist-ai100-2e2025} \\
    
    State/Memory &
    LLM01, LLM04 &
    AML.T0080.000, AML.T0080.001 &
    Persistent memory and thread poisoning
    \cite{Chen2024AgentPoison} \\
    
    Multi-agent &
    LLM01, LLM06 &
    AML.T0051.001 &
    Indirect injection through inter-agent messages; broader propagation
    has no direct one-to-one ATLAS mapping
    \cite{Peign2025,Chen2024,deWitt2025} \\
    
    Supply chain &
    LLM03 &
    AML.T0010 &
    Compromise of models, software, data, or agent tools \\
    
    \bottomrule
    \end{tabularx}
    
    \caption{Mapping of our attack-vector taxonomy to the OWASP Top 10 for
    LLM Applications 2025 \cite{owasp-llm-top10-2025} and MITRE ATLAS
    \cite{mitre_atlas_matrix}. The mappings are conceptual rather than
    one-to-one; in particular, the OWASP taxonomy does not define dedicated
    categories for agent memory or multi-agent communication.}
    \label{tab:mapping_attacks}
\end{table*}

\section{Evidence Synthesis of Attacks and Defenses}
\label{sec:evidence_of_attacks}
    We performed a structured evidence review of peer-reviewed papers and practitioner sources relevant to agentic AI attacks and defenses. We specify the search and inclusion criteria and summarize recurring findings, disagreements, and gaps.
    
 \subsection{Corpus Construction}
We conducted a structured evidence sweep in this paper. 

\emph{Databases/venues.} Queries were issued to ACM Digital Library, IEEE Xplore, arXiv, ACL Anthology, and general search engines (Google Scholar, Bing) using keyword clusters such as ``LLM security tool use,'' ``prompt injection 2024,'' ``RAG poisoning,'' and ``autonomous agent vulnerability.'' We manually reviewed recent proceedings and programs of USENIX Security (2024--2025), IEEE S\&P (2024--2025), CCS (2023--2025), NDSS (2024), SaTML and AISec workshops, and selected industry venues (Black Hat/DEF CON). Standards and industry sources included OWASP GenAI Security, NIST publications, MITRE ATLAS, and technical blogs from model providers and security firms.

\emph{Inclusion/exclusion.} We included works that analyze security of LLM systems with \emph{tools/APIs}, \emph{RAG}, or \emph{multi‑step autonomy} (attacks or defenses). Priority was given to peer‑reviewed papers from 2024 onward; influential preprints were retained when later published or widely cited. Industry/standards documents were included when they offered substantive analysis (e.g., risk taxonomies, evaluative guidance). We excluded purely anecdotal posts and papers on tangential topics (e.g., using LLMs \emph{for} vulnerability discovery rather than attacking agentic systems).

\emph{Screening and coding.} Starting from $\sim$100 candidates (query hits plus snowballing from references), two researchers independently screened titles/abstracts to remove items not about agentic systems, leaving $\sim$40 for full‑text review. We coded each work for evidence type (attack demonstration, measurement, defense evaluation), system setting (open‑source agent, custom pipeline, production integration), and key claims. Items were dropped at full‑text if off‑topic or superseded by stronger, more recent studies.

\emph{Evidence grading.} Each source received an evidence level: \textbf{A} (peer‑reviewed with artifacts or formal proofs), \textbf{B} (peer‑reviewed without artifacts), \textbf{C} (preprint or non‑archival by credible authors), \textbf{D} (industry report/gray literature). We additionally recorded whether a live exploit/demo or quantitative measurements were provided and whether artifacts enable reproducibility.

\emph{Recency and primacy.} For overlapping findings, we cite the most recent or definitive source; when results conflict (e.g., RAG’s safety impact), we report both and attribute differences to setup or metrics.

\emph{Coverage.} The final corpus comprises peer-reviewed papers and a few major preprints (2023--2025), standards and industry documents (OWASP, NIST, MITRE), and a small number of technical blogs and artifacts (e.g., the Help Net Security article summarizing Rezilion’s study \cite{HelpNetSecurity2023}, CETaS prompt-injection guidance \cite{turingIndirectPrompt}).

\subsection{Evidence Synthesis and Key Findings}

Across studies, several patterns recur. First, \emph{LLM+tools implies RCE risk unless proven otherwise}. Empirical analyses show that prompt‑level manipulation readily translates into code execution or unsafe tool calls when schemas, argument handling, or sandboxing are weak; LLMSmith reports 19 RCE vulnerabilities across 11 agent frameworks and successful exploitation of real applications, with subsequent CVE remediation in multiple stacks \cite{Liu2024}. Second, \emph{indirect prompt injection is both practical and difficult to stamp out}. Live demonstrations \cite{Greshake2023} and systematic benchmarking \cite{Yi2025} confirm that models fail to reliably separate retrieved content from instructions; boundary tokens and reminders help but fall short outside white‑box control.

Third, \emph{RAG is not intrinsically safer}. Measurements indicate that retrieval can introduce new failure modes (e.g., unsafe outputs from otherwise benign documents) and that poisoning yields high‑success targeted responses \cite{An2025,Zou2025}. Fourth, \emph{multi‑agent orchestration amplifies risk}. Early testbeds show propagation across agents via malicious messages and partial mitigation via ``vaccination'' at some cost to task performance \cite{Peign2025}.

Defensive results are mixed. Surface defenses (e.g., input sanitization and guardrails) block superficial cases but are bypassable, while structural controls (typed APIs, execution isolation, and human-in-the-loop gating) reduce risk without guaranteeing robustness. Monitoring and anomaly detection remain practical complements when prevention fails \cite{inan2023,zeng2024}.

Across studies, there is consensus that prompt injection is the primary enabling vector and that least-privilege design is essential, while disagreement remains regarding the effectiveness of alignment and policy prompting across setups.

\textbf{Gaps.} Post-exploitation persistence and concealment remain underexplored; fine-grained adversarial perturbations in retrieved text are rare; and common, reproducible testbeds for agent tool misuse are only beginning to appear. Section~\ref{sec:evaluation} proposes evaluation metrics and harnesses to address these gaps. Across surveyed work, tool- and retrieval-level attacks are predominantly supported by grade A–B evidence, while defenses against multi-agent and lifecycle-level threats remain largely at grade C, highlighting a maturity gap.

\section{Evaluation \& Metrics for Agentic AI Security}
\label{sec:evaluation}
We define attacker-aware metrics that capture instruction-following failures, tool/API side effects, and retrieval risks. We show how to compute them from structured execution logs and note pitfalls (non-determinism, evaluator leakage, state carry-over).

\subsection{Defining Key Metrics}
Let $\mathcal{S}$ be test scenarios (benign and adversarial), $\mathcal{A}$ the executed actions (tool/API/file ops), and $\mathbf{1}\{\cdot\}$ an indicator. Logs record timestamps $t$, action types, arguments, costs, policy decisions, and retrieval bundles.

\paragraph{Unsafe Action Rate (UAR)}
The fraction of scenarios that trigger a policy‑violating action: $\mathrm{UAR}=\frac{1}{|\mathcal{S}|}\sum_{s\in\mathcal{S}}\mathbf{1}\{\textsc{violation}(s)\}$.
Variants measure specific classes (e.g., tool‑misuse UAR, leakage UAR). Lower is better.

\paragraph{Policy Adherence Rate (PAR)}
The proportion of executed actions that comply with policy: $\mathrm{PAR}=\frac{\sum_{a\in\mathcal{A}}\mathbf{1}\{\textsc{compliant}(a)\}}{|\mathcal{A}|}$.
Under one‑action‑per‑scenario adversarial testing with crisp labeling, $\mathrm{PAR}\approx 1-\mathrm{UAR}$ only under single-action, adversarial test settings; generally they differ.

\paragraph{Privilege‑Escalation Distance (PED)}
On a causal threat graph $G=(V,E)$ whose nodes are pipeline stages and whose edges encode allowed influences, let $U\subseteq V$ be untrusted inputs and $\Pi\subseteq V$ privileged actions. Define
\[
\mathrm{PED}=\min_{u\in U,\ \pi\in \Pi} \mathrm{dist}_G(u,\pi),
\]
$dist_G(u, \pi)$ the shortest directed path length from untrusted input to privileged action. We treat G as static per deployment and edges as unit-cost unless otherwise specified. Note that higher PED indicates more barriers to escalation.

\paragraph{Time‑to‑Contain (TTC)}
For an incident, with start time $t_{\mathrm{start}}$ and containment time $t_{\mathrm{contain}}$ (automatic kill‑switch or operator action),
\[
\mathrm{TTC}=t_{\mathrm{contain}}-t_{\mathrm{start}}.
\]
Report median and tail percentiles over controlled drills (e.g., induced loops or exfil attempts).

\paragraph{Patch Half‑Life (PHL)}
Let $N(t)$ denote deployments remaining vulnerable to a disclosed issue at time $t$; the half‑life is the smallest $\Delta t$ such that $N(t_0+\Delta t)=\tfrac{1}{2}N(t_0)$. Shorter PHL reflects faster remediation across prompts, policies, and dependencies.

\paragraph{Retrieval Risk Score (RRS)}
For a retrieved bundle $B=\{d_i\}_{i=1}^k$, assign each document a risk $r(d_i)$ from content/source features (e.g., provenance trust, imperative/instructional cues, contamination likelihood, sensitivity). With reliance weights $\alpha_i \ge 0$ (e.g., attention or citation share), normalized such that $\sum_{i=1}^{k}\alpha_i = 1$, $\mathrm{RRS}(B)=\sum_{i=1}^{k}\alpha_i\, r(d_i)$. High RRS can gate autonomy (HITL) or restrict effectors.

\paragraph{Out‑of‑Role Action Rate (OORAR)}
Given a role/tool contract $\mathcal{C}$ specifying allowed actions and parameters:

\[
\mathrm{OORAR}=\frac{\#\{a\in\mathcal{A}:\neg \textsc{allowed}_\mathcal{C}(a)\}}{|\mathcal{A}|}.
\]
where $\textsc{allowed}_\mathcal{C}(a)$ denotes whether action $a$ is permitted by contract $C$. Elevated OORAR suggests drift, mis‑specification, or compromise.

\paragraph{Cost‑Exploit Susceptibility (CES)}
Expected monetary loss under an attacker policy $\pi$ over a fixed horizon $T$:
\[
\mathrm{CES}_T(\pi)=\mathbb{E}_\pi\!\left[\sum_{t=1}^{T} c(a_t)\right],
\]
where $E_\pi[\cdot]$ is the expectation over stochastic agent and environment dynamics,
and $c(\cdot)$ prices API calls, compute, storage, or egress, and $a_t$ denotes the action executed at step $t$. Report worst‑case and rate‑limited variants (e.g., \$ per hour under stress).

\paragraph{Implementation}
A minimal harness randomizes seeds, replays scenarios across $k$ trials, and emits structured traces with unique span IDs. Compute metrics from traces; derive confidence intervals via nonparametric bootstrap. To avoid evaluator leakage, isolate scoring prompts/policies from those that drive the agent; reset state between trials. Handle non-stationarity with time-stratified sampling and periodic scenario recalibration.

  \subsection{Benchmarking Approaches}
We aim to measure security consistently across vectors and paths (P1--P5) with fixed configurations and auditable traces. Each run fixes the model, decoding, agent config, seeds, and tools; traces are exported as JSONL with events, timestamps, tool names, argument/result hashes, and labels. We report UAR/PAR, PED, TTC, and, for RAG, RRS.

\paragraph{Red‑team evaluations}
Automated red‑teaming instantiates families of adversarial prompts and scenarios mapped to known risks (e.g., OWASP GenAI Top‑10), and measure success as UAR. OWASP-aligned tooling (e.g., prompt-generation plugins) adapts to agent contexts for prompt/indirect injection, tool misuse, RAG poisoning, and API abuse \cite{promptfooOWASPPromptfoo}. A balanced suite allocates fixed budgets per vector (e.g., 100 trials each) and reports UAR/PAR with confidence intervals, stratified by capability (code, filesystem, network). Indirect injection suites (e.g., BIPIA) are particularly relevant for browsing/RAG agents \cite{Yi2025}, while prior live exploits inform seed cases \cite{Greshake2023}. For tool safety, include end‑to‑end attempts to elicit code execution or unsafe shell/file actions, building on known failure modes \cite{Liu2024}.

\paragraph{Scenario‑driven testing}
End‑to‑end vignettes exercise full pipelines, including retrieval, planning, tool selection, and side effects. Examples include poisoned README workflows, multi‑agent delegation with a rogue peer, or cloud‑connector abuse. Each scenario specifies preconditions, target effect, privileged boundary, and stop conditions; Outcomes are labeled as success/partial/fail with adjudication rules tied to metrics (e.g., an exfiltration reaching an external sink counts as a UAR violation; a blocked tool call with correct refusal increases PAR). Partial outcomes are conservatively treated as failures for metric computation unless explicitly stated otherwise. Scenario tests often reveal integration faults that unit red-teaming misses.

\paragraph{Continuous evaluation}
Because models, prompts, and policies change frequently, security benchmarks should run in CI/CD with versioned baselines and regression gates (e.g., “no UAR increase $>$ 2\ percentage points on any vector”). Dashboards track trends in UAR/OORAR/TTC and surface drift when new capabilities or tools are enabled. Periodic “fire‑drills” measure Time‑to‑Contain by injecting safe but detectable incidents (e.g., infinite‑loop prompts) and timing automated kill‑switches.

\paragraph{Tooling and harness integration}
Evaluation harnesses should spin up agents in a controlled sandbox, randomize seeds where possible, and emit structured traces (tool calls, arguments, retrieval bundles, costs) for deterministic scoring. Off‑the‑shelf fuzzers and prompt‑search tools can be wrapped as test generators; for RAG, include corpus mutators and index‑time poisoners. Reports must include seeds, agent version, prompt/policy hashes, and tool manifests to support replication.

\paragraph{Non‑stationarity and stochasticity}
Agent outputs vary with decoding and context. To approximate attacker advantage, prefer multiple trials per case and worst‑case reporting (e.g., success if any of $k$ runs succeeds), or use low‑temperature/greedy decoding for lower variance. Confidence intervals (bootstrap) and power analyses should accompany UAR/PAR. Cache is flushed and state reset between trials to prevent inadvertent cross‑contamination.

\paragraph{Evaluator leakage and adjudication}
Never let the system under test grade itself. Use deterministic, side‑channel‑resistant criteria (regex/AST checks on tool requests, filesystem diffs, network egress records) and independent judges when model‑based graders are unavoidable. Sampled human adjudication (e.g., 10--20\% stratified by vector) calibrates and audits automatic labels.

\paragraph{Reproducibility and standardization}
A minimal artifact includes: (i) red‑team prompt suites and RAG corpora with contamination flags; (ii) harness code (Docker) that provisions agents, runs tests, and exports traces; (iii) metric scripts for UAR/PAR/PED/TTC/RRS/ OORAR/CES; and (iv) a reporting template (system version, decoding params, policies, tool list, rate limits). Publishing such suites enables head‑to‑head comparison and longitudinal tracking across agent releases.

\paragraph{Benchmark Properties and Trade-offs.}
Designing agentic security benchmarks involves balancing \emph{completeness}, \emph{realism}, and \emph{feasibility}. Completeness favors broad vector/path coverage (prompt- and indirect-injection, RAG poisoning, tool/API abuse, multi-agent propagation) with adequate trials per cell; realism favors end-to-end scenarios with real stacks, authentic corpora, and noisy conditions; feasibility favors reproducible harnesses and cost/time bounds. In practice, we recommend: (i) a compact, OWASP/ATLAS-mapped core with UAR/PAR/PED reporting; (ii) one or two realism-heavy vignettes per vector family; and (iii) CI-friendly subsets with regression gates (no UAR regressions $>2$\,pp on any vector). This framing clarifies why different organizations may prioritize different slices under resource constraints.

\section{Open Problems \& Research Directions}
\label{sec:open_problems}
Despite rapid progress, securing agentic systems remains an open, multidisciplinary agenda. We outline four research directions with concrete success criteria, evaluation methods, and governance co-design.

\subsection{Formal Methods for Agent Plans and Tool Use}
A promising direction is to model agent plans as programs over typed actions and resources, then verify safety properties (e.g., ``never delete outside a jailed path'' or ``no external email without approval''). Practical approaches include contract‑based design, finite‑state abstractions of planner state, and model checking over a simplified transition system that treats LLM outputs as nondeterministic choices. Useful hybrids constrain the \emph{execution kernel} (schemas, capabilities, effectors) to be deterministic and verifiable, while the LLM remains an untrusted proposer. Key challenges are state explosion from long‑horizon tasks, stochasticity, and partial observability; success criteria include machine‑checkable policies, counterexample traces, and verified enforcement at runtime.

\subsection{Secure Memory and Long‑Term Autonomy}
Persistent memories (histories, scratchpads, vector stores) are durable attack surfaces. Research is needed on provenance‑rich memory architectures with per‑entry trust/confidence scores, decay and re‑validation, and quarantine/rollback semantics. ``Memory antivirus'' routines could scan for known malicious patterns, goal hijacks, or self‑replicating prompts; write paths should be mediated by typed actions and policy. Open problems include detecting subtle, low‑influence poisoning, preserving utility under aggressive sanitization, and maintaining goal integrity over months‑long autonomy.

\subsection{Risk‑Aware Retrieval Beyond Simple Filters}
RAG requires principled selection beyond relevance. Future work should formalize retrieval risk estimation (content, source, time, user entitlements), calibrate it against downstream harms, and couple it to planner decisions (e.g., gating effectors when bundle risk is high). Ideas include ensemble retrievers with outlier suppression, source diversification, cross-source reconciliation to down-weight lone contradictory documents, and integration with fact-checking signals. Methods must balance recall and safety, remain robust to adversarial ``too‑perfect'' matches, and produce interpretable rationales.

\subsection{Continuous Red Teaming and AI‑Augmented Attackers}
Attackers already use generative models to search for jailbreaks and tool‑misuse paths. Defenders need continuous, automated red‑teaming integrated into CI/CD, with generator‑evaluator loops that explore diverse attack families and feed failures back into training, policies, and schemas. Open issues include safe‑by‑design search spaces, avoiding evaluator leakage, and measuring true progress amid non‑stationary models and defenses.

\section{Related Work}
\label{sec:related_work}

Compared to prior overviews that focus on model-centric risks or qualitative guidance for LLM applications, our scope is the agent pipeline with tools, RAG, and autonomy. We use a causal threat graph (PED) and attacker-aware measurements (UAR, PAR, RRS) computed from execution traces. Closest in spirit, Deng et al. survey agent threats~\cite{Deng2025}, and Narajala and Narayan outline an agent-focused threat model~\cite{Narajala2025SecuringAgenticAI}; both emphasize taxonomy and principles. We complement these with a graph that ties vectors to privileged effects and a small measurement harness suitable for CI. TRiSM-oriented reviews~\cite{Raza2025TRiSMAgenticAI} focus on governance and control families; we connect those concerns to specific tool/RAG surfaces and multi-step paths (P1–P5) to aid engineering practice.

Security of language models has begun to receive systematic treatment. Broad surveys of LLM risks (e.g., privacy leakage, adversarial prompting, and mitigation strategies) provide useful background \cite{abdali2024}, but typically center on prompt‑level behaviors in standalone models. Our scope differs by foregrounding \emph{agentic} systems—LLMs coupled with tools/APIs, retrieval pipelines, and autonomous planners—thereby expanding the analysis to planners, effectors, and knowledge infrastructure.

Work on adversarial ML (evasion and poisoning) supplies conceptual foundations for our treatment of RAG threats, which recent studies adapt to generative retrieval settings. In particular, \emph{PoisonedRAG} demonstrates high‑success targeted responses via small corpus perturbations, highlighting indexing‑time and query‑time risks beyond traditional classification \cite{Zou2025}. We place these results in a pipeline model with provenance, access control, and inference-time composition.

Community and industry guidance has cataloged emergent risks and practices. OWASP’s LLM Top‑10 enumerates classes such as prompt injection, insecure output handling, excessive agency, and vector/embedding weaknesses, which we map to our taxonomy and to MITRE ATLAS techniques in Table~\ref{tab:mapping_attacks} \cite{promptfooOWASPPromptfoo}. Industry reports and vendor guidelines surface operational concerns (e.g., supply‑chain exposure and prompt‑injection patterns) that inform our defense‑in‑depth recommendations \emph{(see also Appendix~\ref{sec:defenses_and_mitigation})} \cite{HelpNetSecurity2023,Yi2025}.

Autonomous and multi‑agent security, long studied in distributed systems, is now being revisited in LLM contexts. Early testbeds document cross‑agent propagation, delegation abuse, and partial mitigations (e.g., “vaccination”), connecting classical Byzantine and Sybil considerations to prompt‑mediated channels \cite{Peign2025}. Our SoK integrates these results with tool/RAG vectors to present a unified attack surface for agentic deployments.

On the execution side, decades of work on least privilege, capabilities, and sandboxing are directly applicable but underutilized in LLM toolchains. We emphasize how typed actions, strict schema validation, and containerized effectors translate established systems principles into agent settings \emph{(Appendix~\ref{sec:defenses_and_mitigation})}. Finally, evaluation remains fragmented: while RAG “infusion attack’’ suites and internal red‑teaming have emerged \cite{verma2024}, standardized, attacker‑aware metrics and harnesses are scarce. Our contribution systematizes metrics (UAR, PED, RRS, etc.) and proposes reproducible benchmarking procedures (Section~\ref{sec:evaluation}), complementing prior, largely qualitative guidance.

\section{Conclusion}
\label{sec:conclusion}

Agentic AI—LLMs augmented with tools, retrieval, and autonomy— offers transformative capabilities but introduces a broad and complex attack surface. We systematize this surface via a threat-driven taxonomy: mapping goals, vectors, paths, and preconditions; identifying trust boundaries and TCB; and synthesizing 2023–2025 evidence across prompt/indirect injection, RAG poisoning, tool/API abuse, and multi-agent orchestration. We align risks with OWASP and MITRE ATLAS, define evaluation metrics, and assess defenses across ingestion, inference, planning/acting, and deployment.

Our analysis yields three main conclusions. \textit{First}, no single safeguard suffices; layered defenses—content sanitization, provenance, ACL-aware retrieval, typed schemas, sandboxing, least-privilege credentials, monitoring, and kill-switches—significantly reduce unsafe actions and limit impact. \textit{Second}, security must be measured continuously: attacker-aware metrics (e.g., UAR, PED, RRS, TTC) and CI-integrated red-teaming enable regression control and evidence-based trade-offs between safety and utility. \textit{Third}, cross-disciplinary methods are required, blending formal verification of agent behaviors, hardened retrieval and memory, and policy-as-code governance with technical controls.

Open questions include verifiable autonomy, resilient memory, supply-chain assurance, and attacker-aware benchmarks. Bounding LLM agents with verifiable execution kernels and reproducible evaluations offers a practical path forward. With systematic engineering and red-teaming, these systems can evolve responsibly, reducing a broad attack surface to one manageable for real-world use. Future work should expand empirical testing of these controls in real multi-agent deployments.

\section*{LLM usage considerations}
LLMs were used for editorial assistance. All substantive claims and bibliographic metadata in this revision were reviewed by the authors.

\section*{Acknowledgment}
\label{sec:acknowledgment}
This work was supported in part by the NSERC-CSE Research Community Grants (ALLRP 598786-24), NSERC Discovery Grant (RGPIN-
2019-03995), NSERC Canada Research Chair Grant (CRC-2024-00017), and NSERC CREATE Grant (CREATE 596346-2025) projects 

\bibliographystyle{ACM-Reference-Format}
\bibliography{sample-base}

\appendix
\renewcommand\thesection{\Alph{section}}

\section{Defenses and Mitigations}
\label{sec:defenses_and_mitigation}
We group defenses by where they act: pre-ingestion, inference, agent logic, infrastructure, and monitoring/response. No single control is enough; deployments need defense-in-depth. We note maturity where relevant (established practice vs. tuning-heavy).

\subsection{Pre‑ingestion and Indexing Defenses (Data‑Level)}
These safeguards operate before content reaches the model or agent runtime, and are most relevant to RAG pipelines and any system that ingests untrusted text or documents.

\paragraph{Content sanitization and canonicalization}
Convert HTML/PDF/ Office to a minimal text form; strip or escape markup, scripts, macros, and embedded media. Canonicalization (e.g., DOM→plaintext with simple structure hints) removes many embedded instructions without altering meaning. It does not stop natural-language prompt injection and may drop useful context if over-aggressive. \emph{Maturity:} high; \emph{Limits:} plain-text attacks still apply; possible utility loss.

\paragraph{Active content removal}
Strip or disable executable content (macros, PDF JavaScript) and auto-action links before indexing/display. Run viewers/converters without execution privileges. This blocks system-side triggers unrelated to the LLM. \emph{Maturity:} high; \emph{Limits:} format coverage varies.

\paragraph{Provenance and source trust}
Use allow-lists/signatures for ingestion and tag chunks with trust metadata. At retrieval, down-weight or omit low-trust sources and surface labels to prompts/policies. Gate risky bundles via HITL or effect limits (RRS-based in Section~\ref{sec:evaluation}). \emph{Maturity:} medium; \emph{Limits:} signed $\neq$ true, exceptions must be auditable.

\paragraph{Corpus integrity and change control}
Use append-only/verifiable stores; hash at ingestion; keep tamper-evident update logs; require reviewed pipelines for refreshes. Detects drift and enables rollback but cannot stop first-time poisoning. \emph{Maturity:} medium-high; \emph{Limits:} ops overhead; initial poison still possible.

\paragraph{ACL-aware chunking and retrieval}
Tag embeddings with tenant/role/sensitivity and enforce at query time to prevent cross-principal leakage. Align chunk boundaries with ACLs; avoid mixing principals in a chunk. \emph{Maturity:} medium; \emph{Limits:} perf cost and index bloat.

\paragraph{Poisoning-resilient retrieval}
Limit single-doc influence (caps/  ensembles), re-rank with adversarial filters, and diversify top-$k$ by source/time; flag “too-perfect” matches and near-dups. Where feasible, use robust heuristics that bound impact of a small poisoned fraction. Pair with citation checks/snippet isolation and RRS-based gating. \emph{Maturity:} emerging; \emph{Limits:} tuning-heavy; may reduce recall.

\paragraph{Summary}
Data-level controls lower the chance that malicious content reaches the model and enable risk-aware gating. Combine them with schemas/sandboxes, least privilege, and monitoring for meaningful risk reduction.

\subsection{Retriever and Prompt Processing Defenses (Inference‑Level)}
Once content is admitted to the pipeline, defenses must act at the interface between retrieval and prompt composition to prevent instructions embedded in retrieved text from being treated as executable directives.

\paragraph{Risk‑aware retrieval and ranking}
Retrievers should incorporate a notion of content risk alongside relevance. A practical policy is two‑stage selection: first retrieve by relevance (e.g., top‑$k$), then apply a secondary filter that down‑ranks or excludes items with high risk scores (imperative/instructional cues, unknown provenance, anomalous similarity, recent unvetted sources). This integrates naturally with the Retrieval Risk Score (RRS; Section~\ref{sec:evaluation}): when $\mathrm{RRS}$ exceeds a threshold, the system either withholds risky documents, isolates them (below), or gates effectors (HITL). The trade‑off is potential recall loss if detectors are coarse; high‑assurance deployments prefer conservative thresholds with human fallback.

\paragraph{Snippet isolation and two‑stage use}
Instead of injecting whole documents, isolate minimal supporting snippets and constrain how the model may use them. A two‑stage pattern reduces instruction following: Stage~1 (extract) limits the model to information extraction/quoting from retrieved text; Stage~2 (answer) composes the final response \emph{only} from Stage~1 artifacts. This separation lowers the chance that embedded imperatives are obeyed, and it enables provenance‑aware citation checks. Snippet isolation also supports targeted redaction (e.g., masking secrets) and per‑snippet trust tags that can be surfaced to the planner.

\paragraph{Role segmentation and content labeling}
Preserve role boundaries between system policy, user intent, and retrieved context. Where supported, bind retrieved content to a distinct channel/role and mark as untrusted. Typed schemas (functions/tools with structured args) separate data from instructions so validators can reject smuggled directives in free-text. Role segmentation helps but is not sufficient by itself. Recent structural defenses leveraging formal structured queries, game-theoretic detection sentinels, and preference optimization offer more robust guarantees against injection \cite{Chen2025StruQ, Liu2025DataSentinel, Chen2024SecAlign}.

\paragraph{Constrained decoding and policy-aware generation}
Apply decoding constraints (deny-lists, lexical or pattern filters) to block unsafe outputs and detect high-risk trajectories, such as exfiltration strings or unauthorized tool calls. Policy-aware decoders or post-checks can veto or rewrite results and escalate high-impact actions for human review. These fail-safes catch clear violations but cannot prevent subtle evasions.

\paragraph{Output handling and protocol hygiene}
Treat model outputs as untrusted when rendering or passing to downstream systems. Enforce strict escaping/encoding for UIs and serializers, validate protocol conformance, and quarantine outputs that would break schemas or trigger active content. This addresses “Insecure/Improper Output Handling’’ in the OWASP GenAI Top‑10 and mitigates the agent from becoming a conduit for secondary injection (e.g., UI XSS, log injection) \cite{promptfooOWASPPromptfoo}.

\paragraph{Limitations and integration}
Inference‑level defenses reduce the probability that malicious context is acted upon, but they rely on upstream provenance and downstream effect isolation. In practice, they should be combined with data‑level controls (Appendix~\ref{sec:defenses_and_mitigation}), strict tool schemas and sandboxes (Appendix~\ref{sec:defenses_and_mitigation}), and monitoring/ kill‑switches (Appendix~\ref{sec:defenses_and_mitigation}). Careful tuning balances recall and safety: aggressive filtering improves security metrics (e.g., lowers UAR) at the cost of occasional answer incompleteness.
   
\subsection{Planning and Acting Defenses (Agentic Logic‑Level)}
The core loop that selects and executes actions must treat the LLM as an \emph{untrusted planner} whose proposals are mediated by explicit policies, typed interfaces, and isolation. We outline complementary controls and their trade‑offs.

\paragraph{Capability-based access control (CBAC)}
Expose only required tools and scope each to least privilege (FS, network, IAM). Jail file tools; avoid raw shells; run under a restricted identity/profile with read-only mounts and no default egress. Constrain connector destinations at the tool boundary. Even if plans are subverted, execution stays in-scope.

\paragraph{Typed actions, schema validation, and parameter guards}
Use structured calls and \emph{reject by default} on parse failures. Validate semantics: safe-join paths, enforce domain/protocol allow-lists, cap sizes, and block dangerous tokens/modules. Lax validation led to many exploits~\cite{Liu2024}; strict schemas turn them into refusals. Prefer allow-lists over block-lists.

\paragraph{Sandboxing and effect isolation}
Run tools in containers/VMs with CPU/mem/time quotas, syscall filters, and explicit egress rules. Use per-session ephemeral workspaces. Default to no-network for code; if egress is needed, restrict by destination/protocol and log flows. Cap output size/rate to deter covert exfiltration. Provide fallbacks (e.g., staged artifacts) when policies block effects.

\paragraph{Plan simulation, critics, and effect checks}
Use \emph{plan-then-act}: elicit a plan, check it against policies (tools/paths/scopes), and optionally run a separate “critic’’ model. Approve to execute; otherwise revise or HITL. Static checks handle mechanics; LLM critics catch semantic hazards. Together they lower UAR.

\paragraph{Consequence analysis and self‑reflection}
Before executing high‑impact steps, prompt for anticipated consequences and policy justification (``Is this within policy? Why?''). Use discrepancies between declared safety and detected violations as signals for quarantine or stricter gating. While reflective prompts can be gamed or over‑cautious, their logs aid forensics and operator decisions.

\paragraph{Human-in-the-loop (HITL) gates}
Require approval for sensitive effects (payments, external emails, destructive ops, privilege changes). Escalate based on risk (e.g., high RRS, cross-tenant access). HITL reduces autonomy/latency but reduces the risk of high-impact errors.

\paragraph{Rate limits and quotas}
Bound action rates and cumulative cost (burst and rolling). On threshold, downgrade autonomy, require HITL, or trigger a kill-switch. Mitigates DoC/DoS, buys time for detection, and stabilizes CES.

\paragraph{Secrets management and egress control}
Keep secrets out of the text stream: fetch short-lived, scoped tokens from a vault at call time and inject only at execution. Mask/redact in logs; rotate after incidents. Pair with egress allow-lists and DLP-style checks.

\paragraph{Tamper-evident, externalized logging}
Write append-only, off-box audit logs (tool calls, arg/result hashes, costs, policy decisions, approvals) with integrity protection and synced time. Agents/tools must not modify logs. Immutable traces support IR, regression tests, and TTC/PHL measurement.

\paragraph{Integration and trade-offs}
Combine controls: CBAC/sandboxing bound impact; schemas/guards cut exploits; plan-then-act and HITL catch residuals; quotas/kill-switches contain; logs enable learning. Tune for utility vs. safety: over-restriction raises false positives; permissive settings raise UAR.

\subsection{Infrastructure and Deployment Defenses (Environment‑Level)}
Environment‑level controls constrain what an agent can reach or change regardless of its internal logic. They complement planning/acting defenses by reducing blast radius and improving containment and observability.

\paragraph{Network egress control and micro-segmentation}
Deny egress by default; when needed, restrict to vetted domains/subnets via egress proxies or meshes. Enforce DNS/IP allow-lists, block unneeded internal segments, and log all flows. For browsing/RAG agents, route HTTP(S) through filtering proxies that enforce content-type and domain policy. This limits SSRF-style pivots and exfil paths, improving CES and TTC.

\paragraph{Secure development and supply-chain hygiene}
Pin dependencies, track SBOMs, and patch agent frameworks and connectors promptly. Integrate SCA, static analysis, and fuzzing in CI; gate releases on schema/validator regressions and security checks. Threat-model capability changes (new tools, broader scopes) before enabling.

\paragraph{Policy-as-code and enforcement services}
Encode rules as machine-checkable policies enforced at effect boundaries: DLP on outbound text, protocol checks on tools, encryption for sensitive outputs, and destination allow-lists for connectors. Version and audit policies; run enforcement in independent services so violations are blocked even if the planner is subverted.

\paragraph{Continuous monitoring and anomaly detection}
Operate the agent like a production service: collect telemetry on tool mix, call rates, data volumes, errors, and cost. Alert on deviations (destructive ops spikes, unusual destinations, out-of-role actions) and automatically downgrade autonomy or pause execution on thresholds. Monitoring accelerates IR and supplies traces for UAR, OORAR, CES, and TTC.

\paragraph{Kill-switches and safe interruption}
Provide immediate stop mechanisms: revoke credentials, block egress at the proxy, freeze queues, terminate sandboxes/pods. Make cleanup idempotent and avoid requiring completion of risky actions. Drill periodically to validate latency and logging coverage.

\paragraph{Audits, red teaming, and full-stack scope}
Include agents in security assessments and pen tests. Exercise prompt/indirect injection, tool/schema abuse, RAG poisoning, and connector misuse, and verify hardening of internal APIs, storage, and identity. Track findings to closure and re-test; fold lessons into CI regression suites.

\paragraph{Governance and change management}
Treat prompt/policy updates and new capabilities as production changes: perform risk assessments, update CBAC/sandbox scopes, roll out with feature flags/canaries, and monitor for regressions in security metrics. Record versions and hashes of prompts, policies, and tool manifests for reproducibility and post-incident analysis.

\paragraph{Summary}
Environment-level controls constrain reachability and privileges and improve visibility. Combined with schemas/ CBAC/sandboxes and data-level hygiene, they deliver measurable reductions in unsafe actions and faster containment.

\subsection{Monitoring and Response (Post‑Deployment)}
Post-deployment controls shorten Time-to-Contain (TTC), preserve evidence, and improve Patch Half-Life (PHL). They detect misuse, degrade autonomy under suspicion, and enable rapid, auditable response.

\paragraph{Behavioral anomaly detection}
Profile tool mix, rates, arguments, retrieval risk, and cost; learn per-deployment baselines. Use thresholds (e.g., destructive op spikes, unusual egress) plus sequence models to flag loops, policy drift, and out-of-role actions (OORAR). Include prompts/snippets/proposed effects and confidence in alerts for fast triage.

\paragraph{Risk-adaptive quarantine}
On anomaly or violation, downgrade autonomy: pause effectors, switch to simulate/dry-run, require HITL, and route outputs to moderated channels. Revoke short-lived credentials and restrict egress while preserving session state for investigation.

\paragraph{Forensics-ready telemetry}
Write tamper-evident, append-only logs: prompts; retrieval bundles (IDs/versions/sources); tool calls (name, arg/result hashes); policy decisions; costs; span/trace IDs with synced time. Store conversation history and memory snapshots with minimization and access controls. Hash-chained, off-box storage and clear retention enable reconstruction and UAR/TTC/CES measurement.

\paragraph{Incident response runbook}
Codify detect–triage–contain–eradicate– recover: classify severity; isolate agents/workspaces; block egress; revoke/rotate credentials; snapshot state; notify per obligations. Recover by rolling back corpora/policies, adding regression tests for the triggering vector, and sharing indicators. Drill regularly and record TTC.

\paragraph{Continuous improvement}
Turn incidents and near-misses into tests; update schemas, allow-lists, retrieval policies, and thresholds per root cause; track remediation with PHL. Rotate hidden scenario variants to avoid overfitting and watch utility when tightening controls.

\paragraph{Summary}
Monitoring and response reduce blast radius and dwell time and generate evidence to harden future releases. Coupled with Section~\ref{sec:evaluation} metrics, they provide measurable assurance over time.

\subsection{Governance and Organizational Measures}
Technical controls are effective only within a clear governance frame. 
\emph{Policy.} Define approved uses, data classes, and risk tolerance; keep high-sensitivity actions under HITL until metrics (e.g., low UAR per vector) meet targets. Enforce policy-as-code at boundaries (e.g., DLP filters, connector allow-lists). 
\emph{Training.} Educate developers and operators on agent limits and key threats; encourage hygiene (vet sources, avoid untrusted input) and periodic red-team drills. 
\emph{Audit/accountability.} Maintain append-only logs for prompts, retrievals, and tool calls; assign incident and patching ownership (RACI). 
\emph{Change mgmt.} Treat prompt or policy updates as production changes—risk-assess, restrict scopes, and monitor regression in security metrics.

\begin{figure}[h]
    \centering
    \begin{tikzpicture}
      \draw[fill=bluegray!30, draw=charcoal!50, rounded corners=6pt] (-3.2,0.5) rectangle (4.7,5.6);
      \draw[fill=white, draw=charcoal!50, rounded corners=6pt] (-2.5,0.75) rectangle (4,5);
      \draw[fill=bluegray!20, draw=charcoal!50, rounded corners=6pt] (-2,1) rectangle (3.5,4.2);
      \draw[fill=white, draw=charcoal!50, rounded corners=6pt] (-1.5,1.25) rectangle (3,3.2);
      \draw[fill=bluegray!10, draw=charcoal!50, rounded corners=6pt] (-1,1.5) rectangle (2.5,2.5);
      \node[tcbnode, minimum width=2.2cm, minimum height=0.6cm] at (0.7,2) {Agentic Core};
      \node[font=\scriptsize, align=center] at (0.6,5.3) {%
          \textbf{Input Sanitization \& Provenance}\\[-2pt]
          \begin{tabular}{@{}l@{}}
            • HTML strip \quad • Macro removal \quad • Provenance
          \end{tabular}
        };
      \node[font=\scriptsize, align=center] at (0.6,4.6) {%
          \textbf{Policy/Schema Checks \& Role Separation}\\[-2pt]
          \begin{tabular}{@{}l@{}}
            • Typed actions \quad • Allow lists \quad • Effect checks
          \end{tabular}
        };
      \node[font=\scriptsize, align=center] at (1,3.8) {%
          \textbf{Sandbox \& Capability Scoping}\\[-2pt]
          \begin{tabular}{@{}l@{}}
            • No net \quad • CPU/mem/time caps \quad • Path jail
          \end{tabular}
        };
      \node[font=\scriptsize, align=center] at (0.7,2.9) {%
          \textbf{Monitoring \& Audit}\\[-2pt]
          \begin{tabular}{@{}l@{}}
            • Immutable logs \quad • Anomaly rules \quad • SIEM
          \end{tabular}
        };
      \node[draw=forest, rounded corners=3pt, font=\scriptsize, align=center] 
            at (0.7,0) {Human-in-the-Loop \\ \& Kill-switch};
    \end{tikzpicture}
    \caption{Defense layers around the agentic core with high-level controls.}
    \label{fig:defense_layers}
\end{figure}
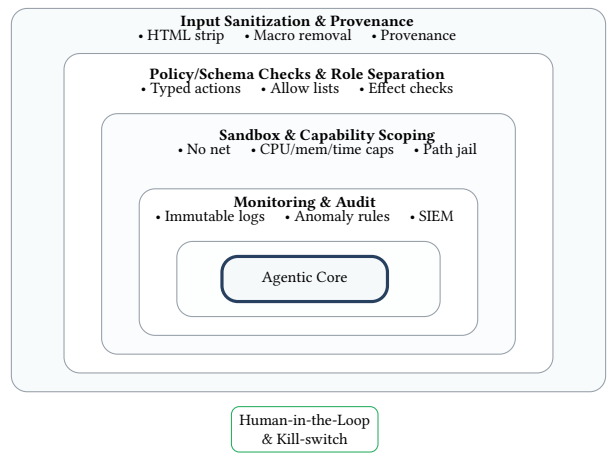

\section{Practitioner Playbook}
\label{sec:practitioner_playbook}
This appendix summarizes concrete steps for securely deploying agentic systems.

\subsection{Pre-Deployment Phase (Design \& Development)}
\textit{Threat modeling.} Apply the system model (Section~\ref{sec:threat_model}); list capabilities (tools/data/connectors), adversaries, and worst-case outcomes. Use these to choose controls and set evaluation targets (e.g., per-vector UAR thresholds).

\textit{Minimal-privilege design.} Provision only needed tools; bind each to least privilege. Avoid unused effectors (e.g., no code/FS tools for pure RAG); add capabilities gradually behind feature flags.

\textit{Prompt/policy.} State refusals; separate user/system/retrieved roles; use EoC markers; never embed secrets. Test drafts against jailbreak/ indirect-injection sets; iterate if targets aren’t met. Pair prompts with typed schemas and effect checks.

\textit{Supply chain.} Pin versions; review advisories/CVEs; vet plugins like code deps; prefer audited/owned plugins; record SBOMs for fast patching.

\textit{Safety nets.} Before launch: strict schemas/guards; sandbox code/file tools (resource + egress limits); ACL-aware retrieval; deterministic output handlers. Add structured, privacy-aware logs (prompts, retrieval IDs/versions, tool calls, policy decisions, costs, trace IDs) to tamper-evident storage.

\textit{Pre-launch testing.} Using Section~\ref{sec:evaluation}, run red-team + scenarios across vectors (PI/indirect, tool/schema abuse, RAG poisoning, connectors). Report UAR/PAR (CIs), PED for critical effects, and TTC from drills; iterate until thresholds are met without undue utility loss.

\textit{Maintenance.} Review threats routinely; update prompts/policies/ models/deps; gate releases on no per-vector UAR regression; track Patch Half-Life. Treat capability changes as production changes (risk, canary, rollback).

\subsection{Deployment and Runtime Phase}
\textit{Secure configuration.} Run agents in isolated containers/VMs (no co-tenanted critical services). Load short-lived secrets from a vault; never hard-code. Bind config to environment (not prompts) and record hashes/versions of prompts, policies, and tool manifests.

\textit{Environment hardening.} Use minimal images; enable host firewalls, read-only mounts, and process/resource limits. In cloud, apply least-privilege IAM so agents can’t provision or reach unused services. Patch base images and kernels promptly.

\textit{Startup checks.} On boot, self-test policy loading, validators, sandbox settings, logging, and egress rules. Emit a signed readiness record (policy hash, tool list, decoding params) to the audit log.

\textit{Continuous monitoring.} Stream telemetry (tool mix, rates, cost, RRS, errors) to SIEM/observability. Alert on anomalies and auto-degrade autonomy (pause effectors, simulate mode) when thresholds trip. Tie health checks to orchestrator actions (quarantine/restart).

\textit{User safeguards.} Add usage guidance in UIs; enforce per-principal rate limits/quotas to deter abuse and bound DoC/DoS. Brief staff on social-engineering risks (e.g., requests to reveal debug modes or prompts).

\textit{HITL operations.} Resource approval queues so HITL decisions are timely. Surface context (tool call, args, retrieved snippets, RRS) to reduce reviewer load and latency.

\textit{Operational reviews.} Periodically review dashboards/logs: blocked violations, alerts, drill TTC, cost outliers. Triage false positives, tune thresholds, and document policy changes.

\textit{Threat intel \& updates.} Track OWASP/NIST/vendor advisories and new vectors; remediate fast. Gate rollouts on no per-vector UAR regression and track Patch Half-Life.

\subsection{Incident Response and Post‑Incident}
\textit{Containment and evidence.} Upon suspected compromise (leakage, unauthorized effect, or high‑risk anomaly), isolate the agent: revoke credentials, block egress, freeze queues, and snapshot sandboxes. Preserve tamper‑evident logs, conversation history, retrieval bundles, and memory snapshots for forensics.

\textit{Notification and compliance.} Classify severity, consult counsel, and notify stakeholders or regulators as required by data‑handling obligations. Maintain an auditable timeline (detection, containment, eradication, recovery) and decision log.

\textit{Root cause and remediation.} Determine the initiating vector (e.g., indirect injection, schema bypass, RAG poison) and close it: tighten schemas/CBAC, adjust prompts/policies, patch dependencies, or remove dangerous tools. Convert the incident into regression tests and red‑team cases; update evaluation suites and dashboards.

\textit{Responsible disclosure.} If the issue reflects a previously unknown vulnerability in third‑party models, frameworks, or services, coordinate responsible disclosure with maintainers to support broad remediation.

\textit{Model and policy adjustments.} Where appropriate, apply fine‑ tuning or instruction updates to improve refusal and role separation; verify there is no undue utility loss. Re‑evaluate with the benchmark harness and publish revised metrics.

\textit{Documentation and exercises.} Record mitigations and lessons learned in living runbooks. Share non‑sensitive practices with the community when feasible. Conduct periodic tabletop exercises that simulate AI‑specific incidents to validate readiness and reduce Time‑to‑Contain.

\subsubsection*{Secure Deployment Checklist}
\begin{itemize}
  \item \textbf{Threat model}: scope tools, assets, attackers, TCB, trust boundaries.
  \item \textbf{Least privilege}: remove unused tools; restrict filesystem access and network egress.
  \item \textbf{Prompts/guards}: enforce role separation; define refusal rules; include jailbreak tests.
  \item \textbf{RAG hygiene}: sanitize ingestion; attach provenance; use ACL-aware chunking.
  \item \textbf{Schemas}: strict JSON; allow‑lists; reject dangerous paths.
  \item \textbf{Sandbox}: no-network code execution; CPU/memory/time caps; syscall filters (e.g., seccomp).
  \item \textbf{Secrets}: short-lived/just-in-time tokens; use a vault; prevent prompt exposure.
  \item \textbf{Observability}: immutable logs; trace/span IDs; PII redaction plan.
  \item \textbf{Gates/quotas}: human-in-the-loop (HITL) for high-impact actions; rate limits; cost caps.
  \item \textbf{Pre‑launch tests}: red-team suite; pass/fail gates; rollback plan.
\end{itemize}

\subsubsection*{Monitoring \& Incident Response Checklist}

\begin{itemize}
  \item \textbf{Signals}: Tool-mix drift; API call spikes; loop detection.
  \item \textbf{Alerts}: remote-code-execution (RCE) patterns; long random strings; off-hours usage.
  \item \textbf{Actions}: auto-quarantine; revoke credentials; kill-switch.
  \item \textbf{Forensics}: snapshot memory; preserve logs; hash-chained integrity.
  \item \textbf{Response}: follow containment runbook; notify stakeholders; rotate keys.
  \item \textbf{Recovery}: patch; run regression tests; post-mortem; update benchmarks.
\end{itemize}

\end{document}